\documentstyle[psfig,aps,prl]{revtex}
\begin{document}
\title{Bures Metrics for Certain
 High-Dimensional Quantum Systems}
\author{Paul B. Slater}
\address{ISBER, University of
California, Santa Barbara, CA 93106-2150\\
e-mail: slater@itp.ucsb.edu,
FAX: (805) 893-2790}

\date{\today}

\draft
\maketitle
\vskip -0.1cm

\begin{abstract}
 H\"ubner's formula for the Bures (statistical distance) metric is
 applied to
both a one-parameter and a two-parameter
 series ($n=2,\ldots,7$) of sets of
 $2^n \times 2^n$ density matrices.
In the {\it doubly}-parameterized  series, the sets
 are comprised of the $n$-fold tensor products --- corresponding to $n$
independent, identical quantum systems --- of the
$2 \times 2$ density matrices with {\it real} entries. The Gaussian
 curvatures of the corresponding Bures metrics are found to be
 constants (${4 \over n}$).
In the second series of $2^n \times 2^n$ density matrices
 analyzed, the {\it singly}-parameterized sets
 are formed --- following an earlier study of Krattenthaler
and Slater --- by averaging with respect to a certain
 Gibbs distribution, the $n$-fold tensor products of the
$2 \times 2$ density matrices with {\it complex} entries.
For $n=100$, we are able to compute
 the Bures {\it distance}
between two arbitrary  (not necessarily neighboring)
 density matrices in this particular series, making
use of certain  eigenvalue formulas of Krattenthaler and Slater,
together with the knowledge that the $2^n \times 2^n$ density matrices in
the series commute.

\end{abstract}


\vspace{-0.1cm}

Some five years ago, H\"ubner \cite{hub1} (cf. \cite {hub2}), in an article
entitled, ``Explicit computation of the Bures distance for density matrices,''
derived a general formula for the Bures or statistical distance \cite{braun}
($d_{B}$) for $n$-dimensional density matrices ($\rho$). It took the form,
\begin{equation} \label{hub1}
d_{B}(\rho,\rho + \mbox{d} \rho)^{2} = \sum_{i=1}^{n} \sum_{j=1}^{n}
{1 \over 2} {| <i| \mbox{d} \rho |j>|^2 \over \lambda_{i} +
\lambda_{j}},
\end{equation}
where $\mbox{d} \rho$ is the incremental
change in $\rho$, and $\lambda_{i}$ is the $i$-th eigenvalue corresponding
 to the
eigenvector $<i|$ of $\rho$.

Dittmann \cite{ditt} has indicated how  formula (\ref{hub1}) can be
reexpressed --- using the Cayley-Hamilton identity --- in terms of certain
 matrix
invariants (thus, obviating any need for the eigenvalues and eigenvectors
themselves). Slater \cite{slat1} has applied Dittmann's formula for the
case  $n=3$ to a set of {\it four}-parameter $3 \times 3$ density matrices.
(More recently still \cite{slatertherm}, we have relatedly examined the
 full {\it eight}-dimensional convex set of
$3 \times 3$, that is spin-1 density matrices.)
In an infinite-dimensional (but three-parameter) setting,
Twamley \cite{twam} has found
 the Bures metric for squeezed thermal states.
Slater \cite{slat2}  utilized these results to find the corresponding
volume element (which turns out to be simply the product of a function of
the squeeze factor and a function of the inverse temperature, the phase
being irrelevant in this regard).
The volume elements of Bures metrics are of particular interest in that --- if
normalizable --- they can be considered to form {\it prior} probability
distributions (for purposes of Bayesian inference)
 over the  associated quantum systems \cite{slat3}.
Twamley \cite{twam} has also suggested that a physical significance can be
attributed to the {\it scalar curvature} of the Bures metric, in providing
a parameterization (coordinate) independent measure of the accuracy of
estimation of a state, given a reference state. (Scalar curvatures of
Riemannian metrics on {\it thermodynamic} state spaces have been a subject of
considerable analysis \cite{brody,rupp}.)

In this communication, we make direct use of the formula of H\"ubner
 (\ref{hub1}) to obtain the restriction of the
Bures metric for
two different series of
 sets of $2^n \times 2^n$ density matrices of interest.
In general, it requires $2^{2 n} -1$ parameters to specify  a
$2^{n} \times 2^{n}$ density
matrix. In light of the consequent
 severe  computational demands entailed, we will limit
 our attention here to certain quite special
density matrices, requiring but one or two parameters for their 
specification. We bring to the reader's attention, however,
 H\"ubner's remark:
 ``The Bures metric is defined on the {\it whole} [emphasis
his] space of density matrices'' \cite[p. 224]{hub2}.

We examine the associated normalized volume elements and the
Gaussian curvatures for these restricted metrics.
We note that in two dimensions --- the framework of our first series
of analyses --- the Gaussian curvature is simply equal to the negative of
one-half of the scalar curvature \cite[p. 184]{lanczos}.
 In one dimension --- the framework of our
second series of analyses --- there is no nontrivial notion of [intrinsic]
curvature.

The starting point for our first series of analyses is the convex set of
 $2 \times 2$ density matrices having {\it real} entries.
Its members ($n=1$) are expressible as
\begin{equation} \label{realmat}
\rho_{real} = {1 \over 2}
\pmatrix{\ 1 + r \cos{\theta}&r \sin{\theta}\cr
      r \sin{\theta}&1-r \cos{\theta}\cr},
\end{equation}
where $(r,\theta)$ are polar coordinates ($0 \leq r \leq 1$,
$0 \leq \theta < 2 \pi$), parameterizing the unit disk.
An application of formula (\ref{hub1}) to the density matrices
(\ref{realmat}) yields the Bures metric,
\begin{equation} \label{1-fold}
d_{B}(\rho_{real},\rho_{real} +\mbox{d} \rho_{real})^2=
g_{r r} \mbox{d}r^{2} +2 g_{r \theta} \mbox{d} r \mbox{d} \theta
+g_{\theta \theta} \mbox{d} \theta^{2},
\end{equation}
where
\begin{equation} \label{Metric1}
g_{r r} = {1 \over 4 (1-r^{2})},
\end{equation}
\begin{equation} \label{Metric2}
g_{r \theta} = 0,
\end{equation}
and
\begin{equation} \label{Metric3}
g_{\theta \theta} = {r^{2} \over 4}.
\end{equation}
We observe that the elements of the metric (\ref{Metric1}) - (\ref{Metric3})
 are {\it independent} of the angular parameter $\theta$.
Normalizing the area   (the form, of course, which ``volume'' takes
 in {\it two} dimensions)
 element ($\sqrt{\det{g}} = \sqrt{g_{r r}
g_{\theta \theta}}$) over the unit disk,
we obtain the prior probability distribution,
\begin{equation} \label {prob1fold}
p(r,\theta) = {r \over 2 \pi \sqrt{1 -r^2}}.
\end{equation}

We have also applied
  H\"ubner's formula (\ref{hub1}) to the $n-$fold tensor product of the density matrix
(\ref{realmat}) with itself --- corresponding to $n$ identical,
independent two-level real quantum
 systems --- for $n=2,\ldots,7$, and obtained, in all these six
 cases, results of the form,
\begin{equation} \label{metric2}
g_{r r}={n \over 4 (1-r^2)},
\end{equation}
\begin{equation} \label{grtheta}
g_{r \theta} = 0,
\end{equation}
and
\begin{equation} \label{gthetatheta}
g_{\theta \theta} = {n r^2 \over 4}.
\end{equation}
(The eigenvalues and eigenvectors --- to be used in (\ref{hub1}) ---  of
 the $n$-fold products are directly derivable through basic rules
from those of $\rho_{real}$ itself.)
Since $g_{r \theta}$ has been found to equal zero, for $n=1,\ldots ,7$,
 the polar coordinates
($r, \theta$) comprise an {\it orthogonal curvilinear}
 (Lam\'e) coordinate system in these instances,
and presumably for all $n$.
(The system is not {\it isothermal} or
 {\it conformal} \cite{thi,tod}, however, in
 that $g_{r r} \neq g_{\theta \theta}$ --- not even for a single value of $r
\in [0,1]$.)

We have, following the lead of Twamley \cite{twam}, computed the
 Gaussian curvature ($K$) --- which equals one-half of the negative of
the scalar curvature \cite[p. 184]{lanczos} --- of the
 Bures metrics reported above.
For our orthogonal (polar) coordinate system, this takes the form \cite[p. 105]
{oprea},
\begin{equation} \label{Gaussiancurv}
K = -{1 \over 2 \sqrt{g_{r r} g_{\theta \theta}}}
 ({\partial \over \partial \theta}
({{\partial \over \partial \theta}
 {g_{r r}} \over \sqrt{g_{r r} g_{\theta \theta}}}) +
{\partial \over \partial r} ({{ {\partial \over \partial r} g_{\theta \theta}}
 \over \sqrt{g_{r r}
g_{\theta \theta}}})),
\end{equation}
For $n=1,\ldots,7$,
 we have that $K={4 \over n}$, that is, the $2^n  \times 2^n$ density
matrices form  spaces of constant positive Gaussian curvature.
(Braunstein and Milburn \cite{braun2} have noted that: ``there is an overall
$\sqrt{N}$ improvement in the precision to which we may determine the
parameter $X$ as we increase the number of identically prepared systems
we can make measurements upon. This is familiar to us as the typical
improvement upon increasing our sample size; here we see that it is a general
limit to how well we can determine a parameter from quantum systems.'')

We have also conducted a more limited analysis of the
 $2 \times 2$ density matrices with {\it complex} entries. These are
parameterizable using spherical coordinates $(r,\theta,\phi)$ in the form,
\begin{equation} \label{complexden}
\rho_{complex} = {1 \over 2}
\pmatrix{\ 1 +r \cos{\theta}& r \sin{\theta} \cos{\phi}-\mbox{i} r \sin{\theta}
\cos{\phi}\cr
r \sin{\theta} \cos{\phi} +
 \mbox{i} r \sin{\theta} \cos{\phi}& 1 - r \cos{\theta}\cr}.
\end{equation}
The associated Bures metric,
\begin{equation} \label{complexmetric}
d_{B}(\rho_{complex},\rho_{complex} + \mbox{d} \rho_{complex})^2 =
g_{rr} \mbox{d} r^2 +2 g_{r \theta} \mbox{d} r \mbox{d} \theta
+g_{\theta \theta} \mbox{d} \theta^2 + 2 g_{r \phi} \mbox{d} r \mbox{d} \phi
+2 g_{\theta \phi} \mbox{d} \theta \mbox{d} \phi + g_{\phi \phi}
\mbox{d} \phi^2,
\end{equation}
for $n=1$, has elements (cf. \cite[formula (15)]{slat}),
 being {\it independent}
of the longitudinal coordinate $\phi$,
\begin{equation} \label{metn2comp1}
g_{r r}= {1 \over 4(1-r^2)}, 
\end{equation}
\begin{equation}
g_{\theta \theta} = {r^2 \over 4},
\end{equation}
\begin{equation} \label{metn2comp3}
g_{\phi \phi} = {r^2 \sin{\theta}^{2} \over 4},
\end{equation}
and
\begin{equation} \label{orthocoord}
g_{r \theta} = g_{r \phi} = g_{\theta \phi} =0 .
\end{equation}
The volume element of this metric
 is normalizable to the prior probability distribution,
\begin{equation} \label{complexprob}
 q(r,\theta,\phi) = {r^2 \sin{\theta} \over \pi^2 \sqrt{1-r^2}},
\end{equation}
over the Bloch sphere \cite{braun2} of two-level quantum systems,
 that is, the unit ball in three-space ($0 \leq r \leq 1$, $ 0 \leq \theta
< \pi$, $ 0 \leq \phi < 2 \pi$).
The (Ricci) scalar curvature  of
 this metric is equal to -24.
We have been able to compute
 the Bures metric for the {\it two}-fold  tensor products
of $\rho_{complex}$ with itself, but the metric elements were given by highly
involved algebraic expressions, which proved difficult to simplify.
 However, by setting $r,\theta,\phi$ to
specific  values a number of times, we obtained results, in all these
 cases, fully
 consistent with the proposition
 that these  elements are, in fact,
 simply
 twice those given by ((\ref{metn2comp1})-(\ref{metn2comp3}). (The scalar
curvature would, then, be equal to -12.)
This would adhere to the pattern noted above
 (for $n=2,\ldots,7$)
 with the $2 \times 2$
{\it real} density matrices. Presumably, there exists a demonstrable theorem
 confirming that this rule holds for all $n$, for both the real and complex
$2 \times 2$ density matrices and, possibly other types of
 density matrices, as well. (The space of $n \times n$ density matrices, for
$n > 2$, ``is not a space of constant curvature and not even a locally
symmetric space, in contrast to what the case of two-dimensional
density matrices might suggest'' \cite{ditt}.)

In our other series of
 analyses, we also apply the formula (\ref{hub1}) of
H\"ubner to a series of $2^n \times 2^n$ density matrices
($n=2,\ldots ,7$).
Rather than the pair of polar coordinates ($r,\theta$), as in the first
instance, these density matrices are parameterized in terms of a {\it single}
 variable ($u$ or alternatively, $\beta$ --- as elaborated below).
The intial (universal quantum coding)
 motivation for studying them was presented in an extended paper
of Krattenthaler and Slater \cite{kratt}. These $2^n \times 2^n$
density matrices were obtained by averaging (over the Bloch sphere of
two-level quantum systems) the $n$-fold tensor products with
 themselves --- corresponding to $n$ independent, identical systems ---
of the $2 \times 2$ {\it complex}
 density matrices (\ref{complexden}). (An analogous [unpublished]
 study has also been
conducted, using the $2 \times 2$ {\it real} density matrices, but the
proofs of certain propositions have turned out --- somewhat
 surprisingly --- to be more problematical, involving an
intricate triple summation, in that [lower-dimensional]
 context.) The averaging
was performed with respect to a
{\it one}-parameter ($u$) family of  probability distributions,
\begin{equation} \label{krattprob}
{\Gamma(5/2-u) r^2 \sin{\theta} \over {\pi}^{3/2} \Gamma (1-u)
(1-r^2)^u}.
\end{equation}
In \cite{slat4} it was argued that this family (\ref{krattprob})
 could be given a
thermodynamic interpretation by using the changes-of-variable,
$u=1-\beta$ and $r =\sqrt{1-e^{-E}}$.
One, then, arrives at a Gibbs distribution of the form,
\begin{equation} \label{Gibbs}
f(E;\beta) ={e^{- \beta E} \over Z(\beta)} \Omega(E),
\end{equation}
where the energy $E$ is taken to be the negative of
 $\log (1-r^2)$, the density-of-states or structure
function, $\Omega(E)$, to be $\sqrt{1-e^{-E}}$, and the partition
 function to be
\begin{equation} \label{partfunc}
Z(\beta) = {\sqrt{\pi} \Gamma (\beta) \over 2 \Gamma(3/2+\beta)}.
\end{equation}
 The natural interpretation of the parameter $\beta$ appears to be that of an
{\it effective polarization temperature}
 \cite{slat4,bross} (cf. \cite{brody2,slateur}).

 Explicit formulas were reported in \cite{kratt} for the eigenvalues
and eigenvectors of the $2^n \times 2^n$ matrices $(\zeta_{n}(\beta))$
 averaged with respect to
(\ref{krattprob}). It was found that there are only $1 +
\lfloor n/2 \rfloor$ distinct eigenvalues.
The eigenvectors constructed in \cite{kratt} formed bases of the
$1+ \lfloor n/2 \rfloor$ subspaces, but were not orthogonalized within the
subspaces (cf. \cite[pp. 426-427]{bied}). The eigenvalues can be expressed as
\begin{equation} \label{eigenkratt}
\lambda_{n,q} = {1 \over 2^n} {\Gamma(3/2+\beta) \Gamma(\beta+q) \Gamma(1+\beta
-q+n) \over \Gamma(\beta) \Gamma(1+\beta +n/2) \Gamma(3/2+\beta+n/2)},
\qquad q=0,1,\ldots, \lfloor n/2 \rfloor
\end{equation}
with respective multiplicities,
\begin{equation} \label{mult}
m_{n,q}={(n-2 d+1)^2 \over (n+1)} {n+1 \choose q}.
\end{equation}
The subspace spanned by the $m_{n,q}$ eigenvectors for the
eigenvalue $\lambda_{n,q}$ corresponds to those explicit spin states
\cite[sec, 7.5.j]{bied} \cite{pauncz} with $q$ spins either ``up''
or ``down'' (and the other $n-q$ spins, of course, the reverse).
The $2^n$-dimensional Hilbert space can be decomposed into the direct sum
of carrier spaces of irreducible representations of $SU(2) \times S_{n}$.
The multiplicities (\ref{mult}) are the dimensions of the corresponding
irreps. The $q$-th subspace consists of the union of $m_{n,q} /(n-2 q+1)$
copies of irreducible representations of $SU(2)$, each of dimension
$(n-2 q+1)$ or, alternatively, of ($n-2 q+1$) copies of irreps of $S_{n}$,
each of dimension $m_{n,q}/(n-2 q+1)$.

The Bures metrics for the six cases ($n=2,\dots ,7$) analyzed, take the simple
form, $g_{\beta \beta} \mbox{d} \beta^2$.
We have found that $g_{\beta \beta}$ equals
\begin{equation} \label{n2}
 {3 \over 4 \beta (2 + \beta) (3+ 2 \beta)^2},\qquad
(n=2)
\end{equation}
\begin{equation} \label{n3}
{ 9 \over 4 \beta (3 + \beta) (3 +2 \beta)^2},
\qquad (n =3),
\end{equation}
\begin{equation} \label{n4}
{9 (145 + 310 \beta +230 \beta^2
+72 \beta^3 + 8 \beta^4) \over
4 \beta (1 +\beta) (3 +\beta) (4 +\beta) (3 +2 \beta)^2 (5+2 \beta)^2},
 \qquad (n = 4)
\end{equation}
\begin{equation} \label{n5}
{15 (185 + 380 \beta + 270 \beta^2
+80 \beta^3 +8 \beta^4) \over 4 \beta (1+\beta) (4 +\beta) (5+\beta) (3 +2 \beta)^2
(5+2 \beta)^2}, \qquad (n=5)
\end{equation}
for $n=6$, the ratio of
\begin{displaymath}
45 (43260 +143640  \beta + 201740 \beta^{2}+
157170 \beta^{3} +
  74361 \beta^4 +
  21864 \beta^5 +
 3896 \beta^6 + 384 \beta^7 +  16 \beta^8)
\end{displaymath}
to
\begin{displaymath}
4 \beta (1 + \beta) (2 + \beta)  (4 +\beta) (5+\beta) (6 +\beta)
(3 +2 \beta)^2 (5 +2 \beta)^2 (7+2 \beta)^2.
\end{displaymath}
and for $n=7$, the ratio of
\begin{displaymath}
63 ( 61950 +  200025 \beta + 273140
\beta^2 +
206472 \beta^3 +  94369 \beta^4
+  26616 \beta^5 +
4504 \beta^6 + 416 \beta^7 
+ 16 \beta^8) 
\end{displaymath}
to
\begin{displaymath}
4 \beta (1 +\beta) (2 +\beta) (5 +\beta) (6+\beta)
(7+\beta) (3 +2 \beta)^2 (5+2 \beta)^2 (7+2 \beta)^2.
\end{displaymath}
For this series of computations, we used MATHEMATICA to obtain fully
orthonormal sets of eigenvectors. In doing so, for $n >4$, computational
considerations required us  to resort to a somewhat
 indirect approach, {\it not} simply
 making use of the Eigensystem command, but rather 
 the NullSpace
command, coupled with our knowledge of the actual eigenvalues. (In all cases,
however, it was necessary to, additionally, employ the GramSchmidt
 command on the
 vectors yielded by the Eigensystem or NullSpace command.)
Since our singly-parameterized matrices for a given $n$ all commute, we could
have relied upon H\"ubner's formula \cite[p. 242]{hub1},
\begin{equation} \label{newhub}
d_{B}(\rho,\rho + \mbox{d} \rho)^{2} =
 \mbox{tr}(\mbox{d} \rho^{1/2})^{2},
\end{equation}
``which is simply the Hilbert-Schmidt metric, not on the space of density
matrices itself, but on the `space of roots of density matrices' rather''
\cite{hub1}.

The numerators of the results ((\ref{n2}), (\ref{n3}))
 for $n=2,3$ are simply  constants. For $n>3$, the roots of the
 numerators also have negative real parts lying
  between -1
 and $-n$.
It is clear from immediate inspection that all the roots of the denominators
for $n=2,\ldots,7$, except for 0, are no greater than -1.
There are, of course, singularities of $g_{\beta \beta}$ at 0 for all these
$n$, and at the other (strictly negative) roots of the denominator.

\begin{figure}
\centerline{\psfig{figure=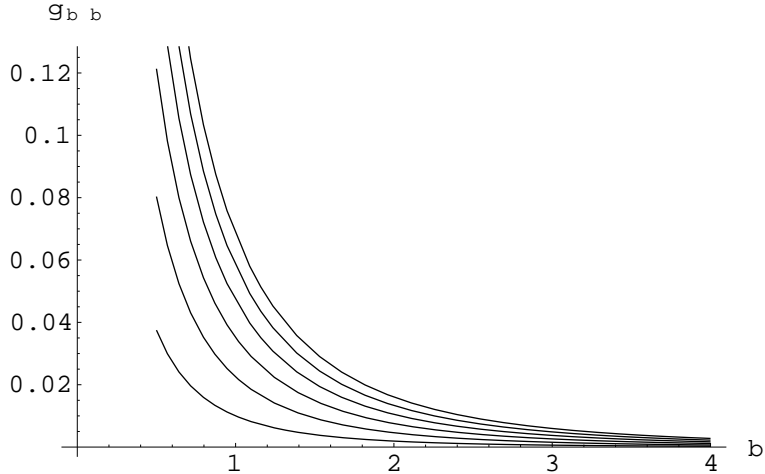}}
\caption{The Bures metric element --- $g_{\beta \beta}$ --- for $n=2,\ldots,7$.
The  dominant curve is that for $n=7$, followed by
that for $n=6,\ldots$}
\label{hubtherm}
\end{figure}

In Fig.~\ref{hubtherm}, we plot $g_{\beta \beta}$ for $n=2,\ldots,7$.
(There is no nontrivial concept of intrinsic curvature for one-dimensional
metrics. In this regard, however, it is of interest to note that in his
study of squeezed thermal states, Twamley \cite{twam} finds  that the 
``curvature
is independent of the 'unitary' parameters $r$ and $\theta$ and only
depends on the `non-unitary' parameter $\beta$.)
The curve for $n=7$ dominates that for $n=6$, which, in turn, dominates that
for
$n=5,\ldots$ All the curves are monotonically decreasing with $\beta$.

The  elements of length ($\sqrt{g_{\beta \beta}}$) can be
normalized over the range $\beta \in [0,\infty]$ by dividing
$\sqrt{g_{\beta \beta}}$ by $\pi /6 \approx$ .523599 ($n$=2),
$\pi /4 \approx .785398$ ($n=3$),  .987405 ($n=4$),   1.1533 ($n=5$), 
1.29428  ($n=6$)
and  1.42688 ($n=7$).

\begin{figure}
\centerline{\psfig{figure=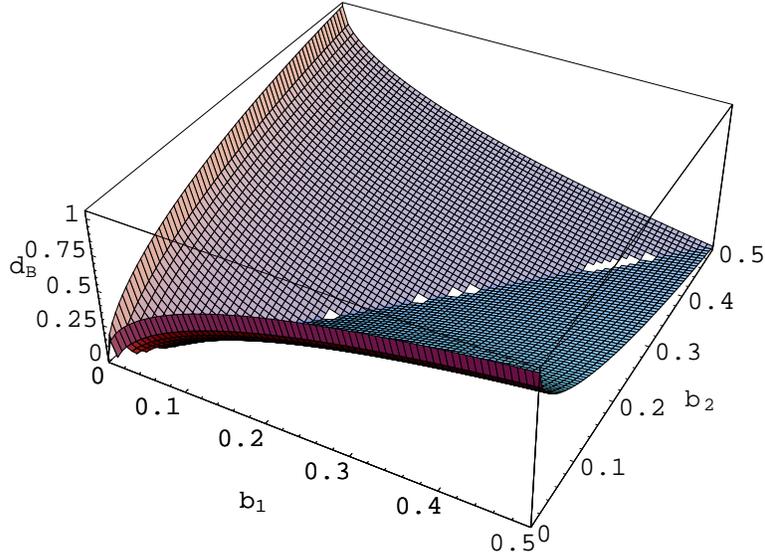}}
\caption{The Bures distance ($d_{B}$)
 between $2^{100} \times 2^{100}$ density matrices,
$\zeta_{100}(\beta_{1})$ and $\zeta_{100}(\beta_{2})$, as a function
of $\beta_{1}$ and $\beta_{2}$.}
\label{figdist}
\end{figure}

From \cite[formula (2.12)]{kratt}, that is, formula (\ref{eigenkratt}) above,
 we know the eigenvalues of the
 averaged matrices
for arbitrary $n$. Since any two averaged matrices $(\zeta_{n}(\beta_{1}),
\zeta_{n}(\beta_{2}))$ for distinct values of
$\beta$  are known from \cite{kratt} to share the same set of 
eigenvectors --- so, $\zeta_{n}(\beta_{1})$ and $\zeta_{n}(\beta_{2})$
necessarily
 commute --- the eigenvalues themselves are all that is required
 to compute the (in general,
{\it nonlocal})
Bures {\it distance} between the density
 matrices. (I thank C. Krattenthaler for pointing this out.) This can be
deduced from
 the general formula for the Bures distance
\cite{hub1},
\begin{equation} \label{dist}
d_{B}(\rho_{1},\rho_{2})^2 = 2 - 2  \mbox{tr} (\rho_{1}^{1/2} \rho_{2}
\rho_{1}^{1/2})^{1/2}.
\end{equation}
Employing (\ref{dist}) for the case $n=2$,
by setting $\rho_{1}=\zeta_{2}(\beta_{1}),
\rho_{2}=\zeta_{2}(\beta_{2})$, we obtain
\begin{equation} \label{buresn=2}
d_{B}(\zeta_{2}(\beta_{1}),\zeta_{2}(\beta_{2}))^2 =
2 -\sqrt{{\beta_{1} \beta_{2} \over (3 +2 \beta_{1}) (3 +2 \beta_{2})}}
-3 \sqrt{{(2+\beta_{1}) (2 +\beta_{2})
 \over (3+2 \beta_{1}) (3 +2 \beta_{2})}}.
\end{equation}
Also, for $n=3$,
\begin{equation}\label{buresn=3}
d_{B}(\zeta_{3}(\beta_{1}),\zeta_{3}(\beta_{2}))^2 =
2- 2 \sqrt{{\beta_{1} \beta_{2} \over (3 +2 \beta_{1}) (3 +2 \beta_{2})}}
-2 \sqrt{{(3 +\beta_{1}) (3 + \beta_{2})
\over (3 +2 \beta_{1}) (3 +2 \beta_{2})}}.
\end{equation}
In Fig.~\ref{figdist}, making a more intensive use of the eigenvalue formula
 (\ref{eigenkratt}),
 we
 plot the Bures distance ($d_{B}$) between
$\zeta_{100}(\beta_{1})$ and $\zeta_{100}(\beta_{2})$.
(Of course, the value of $d_{B}$ along the line $\beta_{1} = \beta_{2}$ is
zero.)
The function shown in this figure is computed as
\begin{equation}
d_{B}(\zeta_{100}(\beta_{1}),\zeta_{100}(\beta_{2}))
 = \sqrt{2 -2 \sum_{q=0}^{\lfloor n/2 \rfloor}  m_{n,q}
 \sqrt{\lambda_{100,q}^{(1)}
 \lambda_{100,q}^{(2)}}}.
\end{equation}
Given the Bures distance,
 we would, then, be able to obtain,
the Bures metric, through the use of the formula \cite[p. 241]{hub1},
\begin{equation}
g_{ij} \mbox{d} \rho^i \mbox{d} \rho^j ={1 \over 2} {\mbox{d}^2 \over
\mbox{d}t^2} [d_{B}(\rho,\rho+t \mbox{d} \rho)^2 \mid_{t=0}.
\end{equation}
Employing such an approach to finding $g_{\beta \beta}$, would  avoid
having to compute the {\it eigenvectors}
of the matrices $\zeta_{n}(\beta)$.
(We, in fact, found the computation of the eigenvectors of the
$256 \times 256$ density matrices, $\zeta_{8}(\beta)$, to be beyond our
resources.)

\begin{figure}
\centerline{\psfig{figure=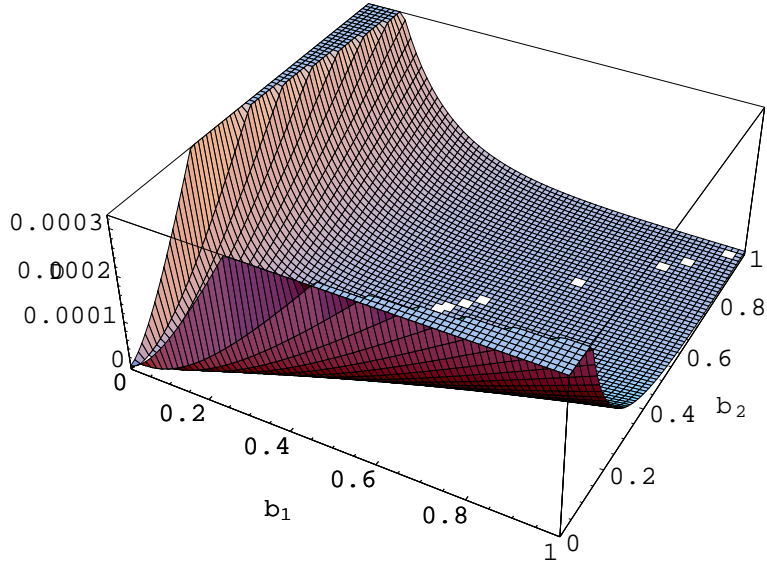}}
\caption{Excess ($\Delta$) over the Bures
 {\it distance} (\ref{buresn=2}) --- for  the case $n =2$ --- of 
the absolute value of the integrated element
 of length  (\ref{intele2}) of the Bures {\it metric}.}
\label{figdiff}
\end{figure}

\begin{figure}
\centerline{\psfig{figure=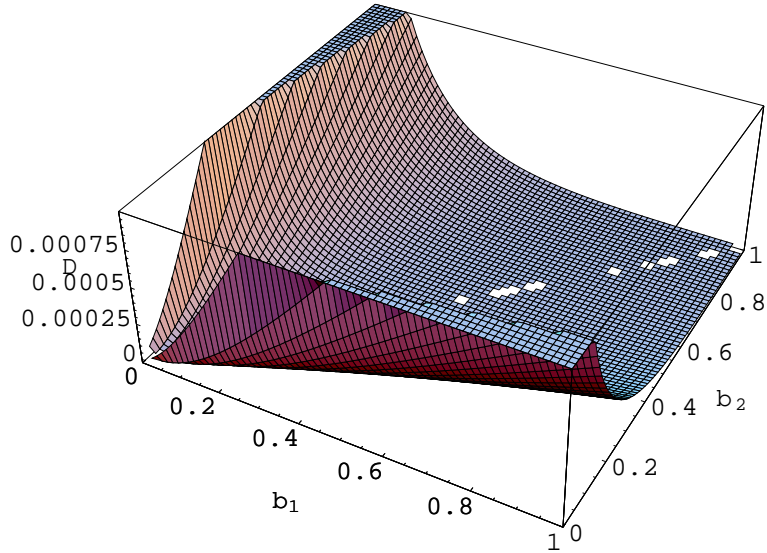}}
\caption{Excess ($\Delta$) over the Bures {\it distance} 
(\ref{buresn=3}) --- for the case $n=3$ --- of
the absolute value of the integrated element of length
 (\ref{intele3}) of the Bures
{\it metric}.}
\label{figdiff3}
\end{figure}

Let us, making use of (\ref{n2}), compare the integrated element of length
\begin{equation} \label{intele2}
\int_{\beta_{1}}^{\beta_{2}} \sqrt{g_{\beta \beta}} \mbox{d} \beta =
\tan^{-1} {\sqrt{\beta_{2}} \over \sqrt{3} \sqrt{\beta_{2} +2}} -
\tan^{-1} {\sqrt{\beta_{1}} \over \sqrt{3} \sqrt{\beta_{1} +2}}, \qquad (n=2)
\end{equation}
with the Bures distance itself (\ref{buresn=2}) between $\zeta_{2}(\beta_{1})$
and $\zeta_{2}(\beta_{2})$.
In Fig.~\ref{figdiff}, we plot the absolute value of this function
(\ref{intele2})
minus the Bures distance --- given by (\ref{buresn=2}).
In Fig.~\ref{figdiff3}, we display the analogous result for the case $n=3$,
making use of (\ref{buresn=3}) and the relation,
\begin{equation} \label{intele3}
\int_{\beta_{1}}^{\beta_{2}} \sqrt{g_{\beta \beta}} \mbox{d} \beta =
\tan^{-1} {\sqrt{\beta{_2}} \over  \sqrt{\beta_{2}+3}} -
\tan^{-1} {\sqrt{\beta_{1}} \over  \sqrt{\beta_{1}+3}}, \qquad (n=3).
\end{equation}
For $n>3$, it appeared it would be necessary
 to employ {\it numerical} integration to evaluate
$\int_{\beta_{1}}^{\beta_{2}} \sqrt{g_{\beta \beta}} \mbox{d} \beta$ to
generate similar figures.

It would be of interest to study the question of whether or not
 the Bures distance between
$\zeta_{n}(\beta_{1})$ and $\zeta_{n}(\beta_{2})$ can be achieved for some
particular  (geodesic)
 path through the {\it unrestricted}  $(2^{2 n}-1)$-dimensional parameter
space of the $2^n \times 2^n$ density matrices (cf. \cite{braun2}).

\acknowledgments

I would like to express appreciation to the Institute for Theoretical
Physics for computational support in this research.

\end{document}